\documentclass[twocolumn,prd,nofootinbib,aps,prl,floats,floatfix,amsmath,amssymb,longbibliography,secnumarabic]{revtex4-1} %
\usepackage[english]{babel}

\usepackage[letterpaper,top=2cm,bottom=2cm,left=3cm,right=3cm,marginparwidth=1.75cm]{geometry}

\usepackage{amsmath}
\usepackage{graphicx}
\usepackage[colorlinks=true,citecolor=red]{hyperref}

\newcommand{\beq}{\begin{equation}}
\newcommand{\eeq}{\end{equation}}

\newcommand{\bea}{\begin{eqnarray}}
\newcommand{\eea}{\end{eqnarray}}

\begin{document}

\title{Comment on ``A try for dark energy in quantum field theory: The vacuum energy of neutrino field''}
\author{James M.\ Cline}
\email{jcline@physics.mcgill.ca}
\affiliation{McGill University Department of Physics \& Trottier Space Institute, 3600 Rue University, Montr\'eal, QC, H3A 2T8, Canada}
\affiliation{Niels Bohr International Academy,The Niels Bohr Institute,
Blegdamsvej 17, DK-2100 Copenhagen Ø, Denmark }

\begin{abstract}
It was recently suggested (\url{https://arxiv.org/pdf/2410.06604}) that the small value of the dark energy of the universe could be explained in terms of the scale of neutrino masses through a simple quantum field theoretic mechanism.  I clarify that the quantity computed is only a threshold correction to the vacuum energy at the low $m_\nu$ scale, and therefore cannot explain the magnitude of the vacuum energy.

\end{abstract}

\maketitle

Recently Ref.\ \cite{Jia:2024kbg} attempted to explain the small observed value of the dark energy density of the Universe in terms of the scale of neutrino masses, starting from the standard expression for their one-loop contribution 
summing the zero-point energies.  It frequently happens that surprising proposals, which sound implausible at the intuitive level, are difficult to immediately debunk because they are expressed in language that is not quite the mainstream parlance (see Refs.\ \cite{Zheng:2024wkp,zheng2024neutrinos,Cline:2024iqp} for an example).  Ref.\ \cite{Jia:2024kbg} is such a case.  The author starts with the quartically divergent expression for the zero-point energies from  a free fermion of mass $m$, and makes subtractions to obtain 
the Coleman-Weinberg formula that, as usual, depends on the renormalization scale,
\beq
    \delta V \sim -{m^4\over 64\pi^2} \ln(m^2/\mu^2)\,.
\eeq 
He  observes that if $m\sim 0.01$\,eV, like for the standard model neutrinos,
and if $\mu$ is slightly larger, then $\delta V$ has the right magnitude to match the observed vacuum energy density. (For previous papers that made a link between the neutrino mass and dark energy scales, see Refs.\ 
\cite{Addazi:2022kjt,
Dey:2022cfg,
Khalifeh:2020bdg,
Gogoi:2020qif,
Salazar-Arias:2019xzx,
NobleChamings:2019ody,
Ennadifi:2017nvl,
Dey:2017wwt,
McKellar:2014hia,
Guendelman:2012vc,
Lambiase:2010ic,
Bhatt:2009wb,
Ando:2009ts,
Bhatt:2008dg,
Antusch:2008hj,
Bamba:2008jq,
Bjaelde:2008yd,
Bhatt:2008hr,
Takahashi:2008epd,
Ichiki:2008rh,
Ichiki:2008st,
Bhatt:2007ah,
Sarkar:2007nt,
Bjaelde:2007ki,
Keum:2007pq,
Gu:2007mi,
Takahashi:2007xp,
Gu:2007ps,
Sahu:2007uh,
Das:2006ht,
Hill:2006hj,
Takahashi:2006be,
Hall:2006br,
Ringwald:2006ks,
Zanzi:2006xr,
Guendelman:2006ji,
Ma:2006mr,
Gu:2005eq,
Zhang:2005ywa,
Li:2005zd,
Fardon:2005wc,
Afshordi:2005ym,
Barbieri:2005gj,
Brookfield:2005td,
Bi:2004ns,
Peccei:2004sz,
Guendelman:2004hk,
Kaplan:2004dq,
Firouzjahi:2022xxb}.)  The author does not explain why one should choose this value of $\mu$, nor why we should ignore all the other particles of the standard model that could contribute similarly.  But it was a good try.

\bibliographystyle{utphys}
\bibliography{sample}

\end{document}